\documentclass{article}
\usepackage{amsmath,amssymb}
\usepackage{latexsym,graphicx,epsfig}

\newcommand{\be}{\begin{equation}}
\newcommand{\ee}{\end{equation}}
\newcommand{\bes}{\begin{subequations}}
\newcommand{\ees}{\end{subequations}}
\newcommand{\bea}{\begin{eqnarray}}
\newcommand{\eea}{\end{eqnarray}}
\newcommand{\bear}{\begin{equation}\begin{array}}
\newcommand{\eear}[1]{\end{array}\label{#1}\end{equation}}
\def\ba{$$\begin{array}}
\def\ea{\end{array}$$}
\def\bra{$\begin{array}}
 \def\era{\end{array}$}

\newcommand{\fr}[2]{\dfrac{{ #1}}{{ #2}}}
\newcommand{\pa}{\partial}
\newcommand{\la}{\langle}
\newcommand{\ra}{\rangle}
\newcommand{\fn}[1]{\footnote{{\sf #1}}}

\newsavebox{\fmbox}

{\end{list}}
\newcounter{enumct}

\begin{document}
\renewcommand{\tilde}{\widetilde}
\renewcommand{\ge}{\geqslant}
\renewcommand{\le}{\leqslant}


\title{ Evolution of
 Universe to the present inert phase}
\author{I. F. Ginzburg, K.A. Kanishev
\\
{\it Sobolev Institute of Mathematics and Novosibirsk State
University},\\
{\it Novosibirsk, Russia}\\
M. Krawczyk, D. Soko\l owska\\
{\it Institute of Theoretical Physics, University of Warsaw},\\
{\it Warsaw, Poland}
}

\maketitle

\begin{abstract}
{
We assume that current state of the Universe can be described by the Inert Doublet Model, containing two scalar doublets, one of which is responsible for EWSB and masses of particles and the second one having no couplings to fermions and being responsible for dark matter.
We consider possible evolutions of the Universe to this state during cooling down of the Universe after inflation. We found that in the past Universe could pass through  phase states  having no  DM candidate. In the evolution via such states  in addition to a possible EWSB phase transition (2-nd order) the Universe sustained one 1-st order phase transition or two phase transitions of the 2-nd order.}

\end{abstract}

\section{Introduction}\label{secIntro}

According to the standard cosmological model about 25\% of the Universe  is made of  Dark Matter (DM). Different candidates for DM particle are now
discussed in the literature. One of the widely discussed  models is the Inert
Doublet Model (IDM) \cite{inert} -- a $Z_2$ symmetric 2HDM with a suitable set of parameters. The model contains one "standard" scalar (Higgs) doublet $\phi_S$, responsible for electroweak symmetry breaking and masses of fermions and gauge bosons as in the Standard Model (SM), and one scalar doublet, $\phi_D$, which doesn't receive  vacuum expectation value (v.e.v.) and doesn't couple to fermions\fn{Our notations are similar to those in the general 2HDM with the change $\phi_1\to\phi_S$, $\phi_2\to\phi_D$.}.

In this model four degrees of freedom of  the Higgs doublet $\phi_S$ are as in the SM:
three Goldstone modes become longitudinal components of the EW gauge bosons and one  mode becomes the Higgs boson (here
denoted as $h_S$).  All the components of the scalar doublet $\phi_D$ are realized
as massive scalar $D$-particles: two charged $D^\pm$ and two neutral
ones $D_H$ and $D_{A}$. By construction, they possess a conserved multiplicative quantum number
and therefore the lightest particle among them can be considered as a candidate for DM particle.

Assuming, as usual, that DM particles are neutral,
we consider such  variant of IDM, in which masses of $D$-particles are
\be
M_{D^\pm}, M_{D_A} \ge M_{D_H} \;\; {\rm {or}} \;\;M_{D^\pm}, M_{D_H}\ge M_{D_A}\,.
\label{chargedheavy}
\ee
Possible masses of these $D$-particles are constrained by the  present accelerator and astrophysical data (see e.g.~\cite{limpap,kra-sok}).

In this paper we assume that the current state of the Universe is described by
IDM. We discuss possible variants of the history of the phase states of  Universe during its cooling down  after inflation. In some respects, this analysis can be considered as particular case of analysis
\cite{GK07}, \cite{GIK09}. We use below some results and notations from \cite{GK07}-\cite{dsthesis}.

\section{The Lagrangian}

In this paper we consider
an electroweak symmetry breaking (EWSB) via the Brout-Englert-Higgs-Kibble (BEHK) mechanism, as described by the Lagrangian
\begin{equation}
{ \cal L}={ \cal L}^{SM}_{ gf } +{ \cal L}_H + {\cal L}_Y(\psi_f,\phi_S) \,, \quad { \cal L}_H=T-V\, .
\label{lagrbas}
\end{equation}
Here, ${\cal L}^{SM}_{gf}$ describes the $SU(2)\times U(1)$ Standard Model
interaction of gauge bosons and fermions, which is independent on the
realization of the BEHK mechanism.
In the considered case the Higgs scalar Lagrangian ${\cal L}_H$ contains the standard kinetic term $T$
and the potential $V$  with two scalar  doublets $\phi_S$ and $\phi_D$.
The ${\cal L}_Y$ describes the Yukawa interaction of fermions $\psi_f$
with only one scalar doublet $\phi_S$, having the same form as in the SM with the change $\phi\to\phi_S$.

\paragraph{Potential.} The potential must be $Z_2$ symmetric in order to describe IDM. Without loss of generality it can be written in the following form
\bes\label{baspot}
\bear{c}
V=-\fr{1}{2}\left[m_{11}^2(\phi_S^\dagger\phi_S)\!+\!
	m_{22}^2(\phi_D^\dagger\phi_D)\right]+	\\[2mm]
+\fr{1}{2}\left[\lambda_1(\phi_S^\dagger\phi_S)^2 \!+\!\lambda_2(\phi_D^\dagger\phi_D)^2\right]+ \lambda_3(\phi_S^\dagger\phi_S)(\phi_D^\dagger\phi_D)+\\[2mm]
	\!+\!\lambda_4(\phi_S^\dagger\phi_D)(\phi_D^\dagger\phi_S) +\fr{\lambda_5}{2}\left[(\phi_S^\dagger\phi_D)^2\!
+\!(\phi_D^\dagger\phi_S)^2\right],
\eear{baspot1}
with all  parameters real  and with additional condition\fn{In the general $Z_2$ symmetric potential  the last term has a
form ~ $\left[\tilde{\lambda}_5(\phi_S^\dagger\phi_D)^2\!+\!{\tilde{\lambda}_5^*}(\phi_D^\dagger\phi_S)^2\right]$.
 The physical content of theory cannot be changed by
  the global phase rotation $\phi_a\to \phi_a e^{i\alpha_a}$ ($a=S,\,D$).
  Starting with an arbitrary complex $\tilde{\lambda}_5=|\tilde{\lambda}_5|e^{i\rho}$
  we select $\alpha_S-\alpha_D=\rho/2+\pi/2$, to get \eqref{baspot}
  with negative $\lambda_5=-|\tilde{\lambda}_5|$.\label{f8}}

\be
\lambda_5<0\,.\label{baspot2}
\ee\ees
The IDM is realized in some regions of parameters of this potential. To study thermal evolution, we will consider also other  possible vacuum states of such potential, at another values of parameters.

To make some equations  shorter, we use the following abbreviations:
 \bear{c}
 \lambda_{345}=\lambda_3+\lambda_4+\lambda_5,\quad
 R=\fr{ \lambda_{345}}{\sqrt{\lambda_1\lambda_2}}.
 \eear{lamnot}

\paragraph{Discrete symmetries.}
This   potential \eqref{baspot} is  invariant under two  discrete
symmetry transformations of a $Z_2$ type:
\bea
S: &	\phi_S \xrightarrow{S} -\phi_S,\quad
	\phi_D \xrightarrow{S} \phi_D,\quad
	SM     \xrightarrow{S} SM,&	\label{ntransf}\\
D: &\phi_S \xrightarrow{D} \phi_S,\quad
	\phi_D \xrightarrow{D} -\phi_D,\quad
	SM     \xrightarrow{D} SM,&
	\label{dtransf}
\eea
where SM denote the SM fermions and gauge bosons.

We  call these transformations {\it $S$-transformation} and {\it
$D$-transformation}, respectively.  In the case when vacuum  has
vanishing vacuum expectation values $\la\phi_{S}\ra=\la\phi_{D}\ra=0$
the mentioned above invariance of $V$ results in the
$D$-parity and $S$-parity conservation in the processes involving only scalars
(or scalars and gauge bosons).
The Yukawa term violates $S$-symmetry
even if
$\la\phi_{S}\ra=\la\phi_{D}\ra=0$, while it respects $D$-symmetry in
any order of perturbation theory.

\paragraph{Positivity constraints.}
To have a  stable vacuum, the potential must be positive at large
quasi--classical values of fields $|\phi_i|$ ({\sl {positivity
constraints}}), for an arbitrary direction in the $(\phi_S,\phi_D)$ plane.
These conditions limit possible values of $\lambda_i$ (see e.g.
\cite{vac-stab}). In terms of parameters \eqref{lamnot}
positivity constraints which are needed in our analysis, can be written as
\bear{c}
\lambda_1>0\,, \quad \lambda_2>0,\quad
R+1>0.
\eear{positivsoft}

\section{Thermal evolution }\label{secevol}

Main goal of this paper is
to consider an evolution of the Universe during its
cooling down to the present inert phase.
For this purpose we consider thermal evolution of the Lagrangian,
following the approach presented in \cite{iv2008,GIK09}.

\paragraph{Potential.} Since the Hubble constant is  small,
we assume a statistical equilibrium at every temperature $T$. In this approximation, at the finite temperature, the ground state of system is given by a minimum of the Gibbs potential
 \be
V_G= Tr\left(V e^{-\hat{H}/T}\right)/Tr\left(
e^{-\hat{H}/T}\right)\,.
 \ee
In the   first nontrivial approximation and high enough temperature the
obtained Gibbs potential has the same form as the basic potential $V$\eqref{baspot},
i.e. as the potential at zero temperature.
The coefficients $\lambda's$ of  the quartic terms  in the potential  $V_G$ and $V$
coincide, while the mass terms vary with temperature $T$, as follows
\bear{c}
m_{11}^2(T)=  m_{11}^2-c_1T^2\,,\quad
m_{22}^2(T) =  m_{22}^2- c_2T^2\,,\\[3mm]
c_1=\fr{3\lambda_1+2\lambda_3+\lambda_4}{12}+\fr{3g^2+g^{\prime 2}}{32}+\fr{g_t^2+g_b^2}{8},\\[2.5mm]
c_2=\fr{3\lambda_2+2\lambda_3+\lambda_4}{12}+\fr{3g^2+g^{\prime 2}}{32}.			 \eear{Tempdep}
Here $g$ and $g^\prime$ are  the EW gauge couplings, $g_t\approx 1$ and $g_b\approx 0.03$ are values of the SM Yukawa
couplings  for $t$ and $b$ quarks, respectively.

Generally each of coefficients $c_1$ and $c_2$ can be either positive or negative.  However,
in virtue of positivity conditions \eqref{positivsoft} their sum is positive,
\be
\label{cesum}
c_2+c_1>0,
\ee
even neglecting  (positive) contributions from gauge bosons W/Z and fermions.

We will show later on that for a realization of the present inert
vacuum with neutral dark matter particle one needs $\lambda_4+\lambda_5<0$ \eqref{nochb}.
Therefore, at $R>0$ we have $\lambda_3>0$. Taking into account
that  $\lambda_5<0$ \eqref{baspot2}, we obtain that $c_1>0$, $c_2>0$.
At $R<0$ there are no constraints on signs of $c_{1,2}$:
\be
   R>0:\quad  c_1>0,\;\;\; c_2>0; \qquad R<0:  \quad {\rm {\,\,arbitrary\,\,\, signs\,\, of\,\,}} c_{1,2}.
\label{signsci}
\ee

\paragraph{Yukawa interaction.} The form of Yukawa interaction and values of Yukawa couplings  don't vary during thermal evolution.

\section{Extrema of the potential}

Following \cite{GK07} we  first consider  extrema of the potential
\eqref{baspot} at arbitrary values of parameters.
The extrema conditions:
\begin{equation}          \label{Eq:min-cond}
\partial V/\partial\phi_i
|_{\phi_i=\langle\phi_i\rangle} =0\,,\qquad \partial
V/\partial\phi_i^\dagger |_{\phi_i=\langle\phi_i\rangle}
=0,\qquad (i=S,\,D)\,
\end{equation}
define the extremum values $\la\phi_{S}\ra$ and $\la\phi_{D} \ra$
of the fields $\phi_{S}$ and $\phi_{D}$, respectively.
The extremum with the lowest
energy  (\emph{the global minimum of the potential}) realizes
\emph{the vacuum state} of the system.
Other extrema are saddle points, maxima  or local minima of the potential.

The most general  solution of
\eqref{Eq:min-cond} can be written in a following form:
\bear{c}
        \langle\phi_S\rangle =\dfrac{1}{\sqrt{2}}\left(\begin{array}{c} 0\\
        v_S\end{array}\right),\quad \langle\phi_D\rangle
        =\dfrac{1}{\sqrt{2}}\left(\begin{array}{c} u \\ v_D
        \end{array}\right),\quad (v^2=v_S^2+|v_D^2|+u^2)
\eear{genvac}
since for each electroweak symmetry violating  extremum (EWv) with $\la \phi_S\ra\neq 0$,
one can choose the $z$ axis in the weak isospin space so that
$ \la\phi_S\ra \sim \begin{pmatrix}0\\v_S\end{pmatrix} $,
with real, nonnegative $v_S$ (choosing a "neutral direction" in the weak isospin space).

\paragraph
{Neutral  extrema.}
The solutions of \eqref{Eq:min-cond} with
$ u=0$  are called neutral extrema, as they respect $U(1)$ symmetry of electromagnetism. For these extrema the  conditions \eqref{Eq:min-cond} can be written as
a  system of two degenerate cubic equations:
\bear{c}
v_S(- m_{11}^2+\lambda_1v_S^2+
\lambda_{345}v_D^2)=0\,,\qquad v_D(-m_{22}^2+\lambda_2v_D^2+
\lambda_{345}v_S^2)=0\,,\\[2mm]
v_S^2\ge 0\,,\qquad v_D^2\ge 0\,.
\eear{nvacCPcons}

This system has four  solutions, one solution defines electroweak symmetric extremum $EWs$
and three solutions define EWSB extrema: inert
extremum $ I_1$, inert-like extremum $ I_2$ and
mixed extremum $M$. Below we list their v.e.v.'s  and extrema energies ${\cal E}_a$:
  \bea
{\pmb {EWs}}:&  v_D=0, \quad v_S=0,\quad\quad \quad\quad {\cal E}_{EWs}=0;&\label{Sol0bas}\\
{\pmb {I_1}}:& v_D=0,\quad v_S^2=v^2=\fr{m_{11}^2}{\lambda_1},\quad
     {\cal E}_{I_1}=-\fr{m_{11}^4}{8\lambda_1};&
     \label{solAbas}\\
{\pmb {I_2}}:& v_S=0,\quad v_D^2=v^2=\fr{m_{22}^2}{\lambda_2},\quad
     {\cal E}_{I_2}=-\fr{m_{22}^4}{8\lambda_2};&
     \label{solBbas}\\
{\pmb M}:&
    \begin{array}{c}
      v_S^2=\fr{m_{11}^2\lambda_2-\lambda_{345}m_{22}^2}{\lambda_1\lambda_2-\lambda_{345}^2},\quad
      v_D^2=\fr{m_{22}^2\lambda_1-\lambda_{345}m_{11}^2}{\lambda_1\lambda_2-\lambda_{345}^2};\\[4mm]
      {\cal E}_{M}=-\fr{m_{11}^4\lambda_2-2\lambda_{345}m_{11}^2m_{22}^2+m_{22}^4\lambda_1}
      {8(\lambda_1\lambda_2-\lambda_{345}^2)}.
    \end{array}&
\label{Nextr1}
\eea
Some of the equations \eqref{solAbas}-\eqref{Nextr1} can give also negative values
of  $v_S^2$ or $v_D^2$, in contradiction with  basic condition for the extremum
\eqref{nvacCPcons}. In  such case the extremum, described by corresponding equations,
is absent.

The energy differences between  $I_{1,2}$ and $M$ extrema are as follows:
\be
{ \cal E}_{I_1}-{\cal E}_{M}
       =\fr{\left(m_{11}^2\lambda_{345}-m_{22}^2\lambda_1\right)^2}{
       8\lambda_1^2\lambda_2(1-R^2)};\quad { \cal E}_{I_2}-{\cal E}_{M}
       =\fr{\left(m_{22}^2\lambda_{345}-m_{11}^2\lambda_2\right)^2}{
       8\lambda_1\lambda_2^2(1-R^2)}.
\label{compE}
\ee

\paragraph{Charge breaking  extremum.} For  $u\ne0$ the  extremum
violates not only EW symmetry but also the $U(1)$ electromagnetic symmetry,
leading to the electric charge non-conservation.
According to general analysis in \cite{lorenzo,vac-others,Barroso:2007rr,GK07,dsthesis}
this extremum can realize vacuum state only if:
\begin{equation}
\lambda_4+\lambda_5 >0.
 \end{equation}
We will see later on that at this condition the DM particle
can not be neutral, that contradicts  \eqref{chargedheavy}.

\section{ Vacuum states}\label{secvacst}
Below we describe briefly properties of neutral extrema,  listed in the previous section,
provided that they are realized as  true vacua.

\subsection{Electroweak symmetric  vacuum $EWs$}

The electroweak symmetric extremum with $\la\phi_S\ra=\la\phi_D\ra=0$ exists for all values of parameters of the potential~\eqref{baspot}. It
respects the  $D$ and $S$-symmetries  of the potential.
This extremum is a minimum, realizing vacuum state, at
\begin{equation}
 m_{11}^2<0,\qquad m_{22}^2<0.
\end{equation}
In this case, gauge bosons and fermions are massless, while scalar doublets  $\phi_S$ and $\phi_D$ have masses equal to $|m_{11}|$ and $|m_{22}|$, respectively.

\subsection{Inert vacuum $I_1$}\label{secinert}

In the case when $I_1$ extremum realizes vacuum,  the Inert Doublet Model describes reality.
The standard field decomposition near $I_1$ extremum has a form
\be
\phi_S=\begin{pmatrix}G^+\\ \fr{v+h_S+iG}{\sqrt{2}}\end{pmatrix}\,,\qquad \phi_D=
\begin{pmatrix}D^+\\ \fr{D_H+iD_A}{\sqrt{2}}\end{pmatrix}\,,
\label{decompA}
\ee
where $G^\pm$ and $G$ are Goldstone modes, while $h_S$ and $D=\,D_H,\,D_A,
D^\pm$ are scalar particles.  Here the Higgs particle $h_S$ interacts
with the fermions and gauge bosons just as the Higgs boson in the SM.

\paragraph{Symmetry properties.} The inert vacuum state violates the $S$-symmetry
\eqref{ntransf}. However, this state is invariant under the $D$-transformation
\eqref{dtransf} just as the whole basic Lagrangian \eqref{lagrbas}. Therefore the  $D$-parity is conserved, and  due to this fact
the lightest $D$-particle is stable, being a good  DM candidate.

\paragraph{Allowed region of parameters.}
For the \emph{inert extremum} to exists it is necessary that $m_{11}^2>0$ \eqref{solAbas}. In accordance with
\eqref{solAbas} and \eqref{solBbas},  the extremum $I_1$ can be a \emph{vacuum} only
if $m_{11}^2/\sqrt{\lambda_1}>m_{22}^2/\sqrt{\lambda_2}$. Additional condition arises from a comparison of $I_1$ and $M$ extrema. In virtue of \eqref{compE} at
$1-R^2<0$ the extremum $M$ can exist but its energy is larger than energy of $I_1$ extremum -- so that the extremum $I_1$  realizes vacuum.
At $1-R^2>0$ the inert extremum still can be a vacuum,
in the case when the mixed extremum does not exist,
i.e. if at least one of  quantities $v_S^2$, $v_D^2$ defined by
eq.~\eqref{Nextr1} is negative.

Note, that due to the positivity constraint  $1+R>0$  \eqref{positivsoft}  in the case when $1-R^2<0$
we have $R>1 $. For the opposite case, with $1-R^2>0$, the quantity $R$ can be either positive or negative.

\paragraph{Particle properties.} The quadratic part of the potential
written in terms of physical fields $h_S,\,D_H,\,D_A$ and
$D^\pm$  \eqref{decompA} gives the following masses of scalars:
\bear{c}
M_{h_S}^2=\lambda_1v^2= m_{11}^2\,,\qquad M_{D^\pm}^2=\fr{\lambda_3 v^2-m_{22}^2}{2}\,,\\[3mm]
M_{D_A}^2=M_{D^\pm}^2+\fr{\lambda_4-\lambda_5}{2}v^2\,,\qquad M_{D_H}^2=
M_{D^\pm}^2+\fr{\lambda_4+\lambda_5}{2}v^2\,.
\eear{massesA}
The requirement that lightest $D$-particle is a neutral one
\eqref{chargedheavy} results in the  condition
\begin{equation}
\lambda_4 + \lambda_5 <0. \label{nochb}
\end{equation}
Since
$\lambda_5<0$ \eqref{baspot2}  the "scalar" $D_H$ is lighter than
"pseudoscalar"\fn{Note that the rephasing transformation $\phi_1\to \phi_1$, $\phi_2\to i\phi_2$, changing sign of $\lambda_5$, results in change $D_H\leftrightarrow D_A$ in $I_1$ state.} $D_A$.

As in the standard 2HDM, scalars $D_H$ and $D_A$ have opposite $P$-parities but since
they don't couple to fermions, there is no way to assign to  them
a definite value of $P$-parity. However, their relative parity does matter
and for example, vertex $ZD_HD_A$ is allowed while vertices $ZD_HD_H$ and  $ZD_AD_A$ are forbidden.
According to our basic assumption on the Yukawa interaction
\eqref{lagrbas} $D$-particles don't interact with fermions.
Neither there are interactions of $D$-particles with gauge bosons
$V$ of the type $D_iV_1V_2$.

\subsection{Inert-like vacuum $I_2$}

\emph{The inert-like vacuum} $I_2$  is "mirror-symmetric" to the inert vacuum
$I_1$, compare \eqref{solAbas} and \eqref{solBbas}. The interaction among scalars and between scalars and gauge bosons  are
mirror-symmetric as well, so the only
difference between $I_2$ and $I_1$ arises from the Yukawa interaction.

Main formulae for this state are similar to those for the vacuum $I_1$ with
obvious replacements.  The corresponding field decomposition is given by
\be
\phi_S=\begin{pmatrix}S^+\\ \fr{S_H+iS_A}{\sqrt{2}}\end{pmatrix}\,,\qquad \phi_D=
\begin{pmatrix}G^+\\ \fr{v+h_D+iG}{\sqrt{2}}\end{pmatrix}\,,
\label{decompB}
  \ee
with  one Higgs particle $h_D$ and  four $S$-particles: $S_H,\,S_A,\,S^\pm$.

\paragraph{Symmetry properties.} The inert-like vacuum $I_2$  violates
 $D$-symmetry \eqref{dtransf}. This state as well as the Higgs potential  is
invariant under the $S$-transformation \eqref{ntransf}.
However, in contrast to the inert vacuum, here $S$-parity is not conserved
by the whole Lagrangian because of the form of Yukawa interaction.

\paragraph{Allowed regions of parameters.}
The  \emph{inert-like extremum} exists for $m_{22}^2>0$.  In order to have an
\emph{inert-like vacuum} it is necessary that
$m_{11}^2/\sqrt{\lambda_1}<m_{22}^2/\sqrt{\lambda_2}$. For $1-R^2<0$
there are no additional demands. If $1-R^2>0$
inert-like extremum can be a vacuum only if at least one of quantities $v_S^2$, $v_D^2$, defined by eq.~\eqref{Nextr1}, appears to be negative.  Both these conditions are similar to those for the inert vacuum $I_1$.

\paragraph{Particle properties.} The masses of the Higgs boson $h_D$  and $S$-scalars
are given by (cf. \eqref{massesA})
\bear{c}
   M_{h_D}^2=\lambda_2v^2= m_{22}^2\,,\qquad M_{S^\pm}^2=\fr{\lambda_3 v^2-m_{11}^2}{2}\,,\\[3mm]
   M_{S_A}^2=M_{S^\pm}^2+\fr{\lambda_4-\lambda_5}{2}v^2\,,\qquad M_{S_H}^2=
   M_{S^\pm}^2+\fr{\lambda_4+\lambda_5}{2}v^2\,.
\eear{massesB}

The Higgs boson $h_D$ couples to gauge bosons just as the Higgs boson of the SM, however
it does not couple to fermions at the tree level. The $S$-scalars do interact with fermions.
Therefore, here there are  no candidates for dark matter particles. That is the reason to
call this vacuum {\it inert-like vacuum}.

Note that all fermions, by definition interacting only with $\phi_S$ with vanishing v.e.v.
 $\la\phi_S\ra =0$,  are massless. (Small mass can appear only as a loop effect.) In such vacuum state
particles form roughly uniform plasma with massless fermions and heavy gauge bosons and scalars.

\subsection{Mixed vacuum $M$}

The mixed extremum\fn{Sometimes called  a\emph{ normal extremum} $N$, see e.g. \cite{Barroso:2007rr}}  $M$  violates both $D$- and $S$-symmetries, i.e. the full $Z_2$ symmetry of
the potential. In this vacuum we have massive fermions and no candidates for DM particle,   like
in the SM. The decomposition around the mixed vacuum looks as follows:
\be
\phi_S=\begin{pmatrix}\rho_S^+\\ \fr{v_S+\rho_S+i\chi_S}{\sqrt{2}}\end{pmatrix}\,,\quad \phi_D=
\begin{pmatrix}\rho_D^+\\ \fr{v_D+\rho_D+i\chi_D}{\sqrt{2}}\end{pmatrix},
\label{decompM}
  \ee
where the $\rho_S^+$ and $ \rho_D^+$ lead to two orthogonal combinations $G^+$ and $H^+$, while
 $\rho_S$ and  $\rho_D$ ($\chi_S$ and  $\chi_D$)
-- to two orthogonal combinations $h$ and $H$ ($G$ and $A$), respectively. There are here five Higgs bosons - two charged $H^\pm$ and three neutral ones: the CP-even $h$ and $H$ and CP-odd  $A$.

\paragraph{Allowed regions of parameters.}
In accordance with \eqref{Nextr1} and \eqref{compE}  the mixed extremum is global minimum of potential, i.e. \emph{vacuum}, if and only if the following conditions hold: $v_S^2>0, \,\,v_D^2>0$ and $1-R^2 >0$.
For v.e.v.'s squared given by eqs.~\eqref{Nextr1} the latter conditions
can be transformed to the relations between  mass
parameters $m_{11}^2$ and $m_{22}^2$:
\bear{lc}
{\rm at}\;\;\;\;\;\; 1>R>0: & 0<R\fr{m_{11}^
2}{\sqrt{\lambda_1}}<\fr{m_{22}^2}{\sqrt{\lambda_2}}<\fr{m_{11}^2}{R\sqrt{\lambda_1}}\, ;\\[2mm]
{\rm at}\;\;\;\; 0>R>-1:& \fr{m_{22}^2}{\sqrt{\lambda_2}}>R\fr{m_{11}^2}{\sqrt{\lambda_1}},
\quad \fr{m_{22}^2}{\sqrt{\lambda_2}}>\fr{m_{11}^2}{R\sqrt{\lambda_1}}\,.
\eear{Ccond2}

\paragraph{Particle properties.}  Masses of scalars are as follows (see, e.g.
  \cite{GK05,GK07})
\be
 M_{H^\pm}^2=-\fr{\lambda_4+\lambda_5}{2}v^2\,,\quad
 M_A^2=-v^2\lambda_5,\quad \left(v^2=v_S^2+v_D^2\right). \label{chneitr}
\ee
The neutral CP-even  mass matrix is equal to
\be
 {\cal{M}}=\begin{pmatrix}\lambda_1v_S^2&\lambda_{345}v_Sv_D\\
                             \lambda_{345}v_Sv_D&\lambda_2v_D^2\end{pmatrix}\,.
\label{massmatrixCV}
\ee

Note, that the extremum can be minimum only if both diagonal elements of mass matrix
and its determinant are positive, i.~e. $\lambda_1\lambda_2v_S^2v_D^2(1-R^2)>0$,
in agreement with the above mentioned conditions.
It means also that in the case if mixed extremum is minimum, it is global minimum -- vacuum.

The  mass matrix \eqref{massmatrixCV} gives masses of the neutral CP-even Higgs bosons:
\be
 M_{h,H}^2=\fr{\lambda_1v_S^2+\lambda_2v_D^2\pm
 \sqrt{(\lambda_1v_S^2+\lambda_2v_D^2)^2-4\det{\cal M}}}{2}\,,\label{massesC}
\ee
with sign $+$ for the $H$ and sign $-$ for $h$.

Couplings of the physical Higgs bosons to fermions and gauge bosons have
standard forms as for the  2HDM, with the Model I Yukawa interaction.

\section{Evolution of phase states of the  Universe}\label{secevol2}

In this section we consider possible phase history of the Universe, leading to the inert vacuum $I_1$ today, using the thermal evolution described in sec. 3.

To summarize properties of different vacua of the $Z_2$-symmetric potential and to classify all possible ways of evolution of the Universe  we will use phase diagrams
in the $(\mu_1(T),\,\mu_2(T))$ plane,  where
\be
\mu_1(T)= m_{11}^2(T)/\sqrt{\lambda_1},\qquad \mu_2(T)= m_{22}^2(T)/\sqrt{\lambda_2}\,\,\,.
\ee

Let us remind (sect.~\ref{secevol}) that in our approximation during  cooling down of Universe parameters $\lambda_i$ are fixed, while mass parameters $m_{ii}^2$ vary.
These  variations result in modification of vacuum state and a possible change of its nature. Possible types of  evolution depend on value of parameter $R$ \eqref{lamnot} and  are depicted in the figures~\ref{muplot1},~\ref{muplot2} and~\ref{muplot3}. The possible current states of Universe are represented  in these figures by small black dots  $P= (\mu_1,\,\mu_2)$ \fn{In  subsequent discussion we will distinguish  present day values of parameters $\mu_i\equiv\mu_i(0)$ and their values $\mu_i(T)$ at some temperature $T$.}.
Since currently we are  in the inert phase, we have $\mu_1>0$ for each point $P$ (see sect.~\ref{secinert}).
The parameter $\mu_2$ can be both positive (points $P1$ and $P3$) and negative (points $P2$, $P4$ and $P5$).

In accordance with \eqref{Tempdep} a particular evolution leading to a given  physical vacuum state $P$ is represented by a  ray, that ends at a point $P$. Arrows on these rays are directed towards a  growth of time (decreasing of temperature).
The direction of the ray is determined by parameters (cf. \eqref{Tempdep})
\be
\tilde c_1=c_1/\sqrt{\lambda_1},
\qquad \tilde c_2=c_2/\sqrt{\lambda_2}, \qquad \tilde c=\tilde c_2/\tilde c_1.\label{ctilde}
\ee

For different possible positions of today's point $P$ we consider typical  evolutions for different possible values of parameter $\tilde{c}$.
In figures below  all representative rays are shown; they are labeled by two numbers, with the first one corresponding to the label of the final point $P$.

\subsection{The case $\pmb{R>1}$}\label{secevola}

\begin{figure}[htb]
\centering
\includegraphics[width=0.5\textwidth]{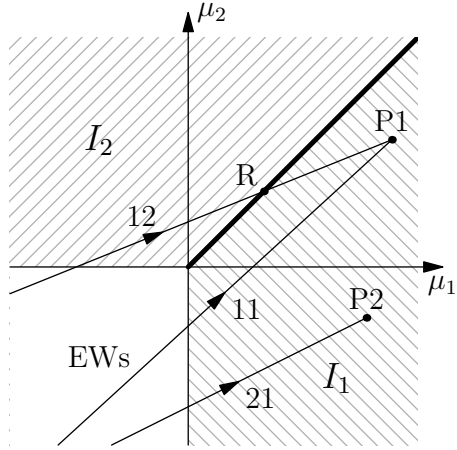}
\label{muplot1}
\caption{Phase diagram for $R>1$ case. }
\end{figure}

Phase diagram for this case is presented in Fig.~\ref{muplot1}. It contains one quadrant with $EWs$ phase and two  sectors, describing the $I_1$ and $I_2$
phases. These two phases $I_1,I_2$  are separated by {\it the phase transition line  $\mu_1=\mu_2$} (thick black line). Two  typical positions of today's state are represented by points $P1$ ($\mu_2>0$) and $P2$ ($\mu_2<0$). Since  (according to \eqref{signsci}), both $\tilde c_1, \tilde c_2>0$ ($\tilde c>0$),
all possible   phase evolutions are represented by  rays 11 and 12 for the today's point $P1$ and by  a  ray 21 which leads to the today's point $P2$.

\paragraph{Ray 11:\hspace{2mm}$\pmb{ \tilde c > \mu_{2}/\mu_{1}>0 } $.}
The Universe started from the EWs state and after the second-order EWSB transition at $m_{11}^2(T)=0$, i.e. at the temperature
\be
 T_{EWs,1}=\sqrt{m_{11}^2/c_1}=\sqrt{\mu_1/\tilde c_1}\,,\label{TEWSBI_1}
\ee
has entered to the present inert phase $I_1$.

\paragraph{Ray 12:\hspace{2mm}$\pmb{ 0< \tilde c < \mu_{2}/\mu_{1} } $. }
The Universe started from the $EWs$ state. Then it went through the EWSB second-order
phase transition into the inert-like phase $I_2$ at $m_{22}^2(T)=0$, i.e. at the temperature equal to
\be
T_{EWs,2}=\sqrt{m_{22}^2/c_2}=\sqrt{\mu_2/\tilde c_2}.\label{TEWSBI_2}
\ee
The next transition is the  phase transition from
the  inert-like phase $I_2$ into the today's inert phase $I_1$ at the point $R$, where $\mu_2(T)=\mu_1(T)$, i.e.  at the temperature
\be
T_{2,1}=\sqrt{ \fr{\mu_1-\mu_2}{\tilde c_1 -\tilde c_2}}\,. \label{TI1I2}
 \ee
That is the first-order phase transition with the latent heat given by
\bear{c}
 Q_{I_2\to I_1}=\left. T\fr{\pa{\cal E}_{I_2}}{\pa T}-T\fr{\pa{\cal E}_{I_1}}{\pa T}
\right|_{\mu_2(T)\to\mu_1(T)}=({\mu_{2}\tilde c_1-\mu_1\tilde c_2})T^2_{2,1}/4\,.
\eear{heat}

\paragraph{Ray 21:\hspace{2mm} $\pmb{\mu_{2}<0}$.}
The Universe started from $EWs$ state and after the second-order EWSB transition
at the temperature \eqref{TEWSBI_1} has entered  the  today's phase.

\subsection{The case $\pmb{1>R>0}$}
\label{secevolb}

\begin{figure}[htb]
\centering
\includegraphics[width=0.5\textwidth]{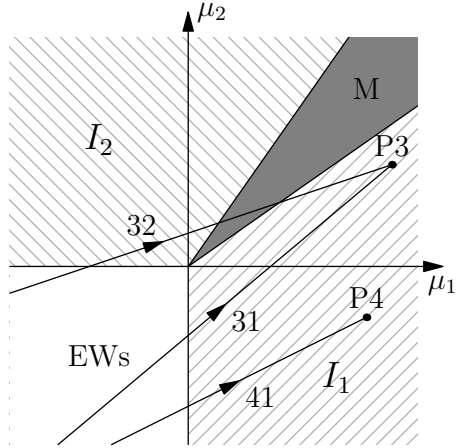}
\label{muplot2}
\caption{Phase diagram for $1>R>0$ case. }
\end{figure}

The phase diagram in Fig.~\ref{muplot2} is modified in comparison with the Fig.~\ref{muplot1},
as according to  \eqref{Ccond2} in the upper right quadrant of the considered $(\mu_1,\mu_2)$ plane
the new sector -- the mixed phase $M$ -- occurs in the region
\be
0< R\mu_1<\mu_2< \mu_1 /R.
\label{Mphaseb}
\ee
As before, since $R>0$ both $c_{1,2}>0$ and consequently $\tilde{c}>0$.

Since currently we are in the inert vacuum, the possible  today's states are of type of points $P3$ and $P4$, for which
\be
 \mu_2< R\mu_1.\label{inmuM}
\ee
All possible phase evolutions are represented by three rays in
Fig.~\ref{muplot2}, with rays 31 and 32 having the today's endpoint $P3$ while the ray 41 is pointing  $P4$.

For the { rays 31} and  { 41},  phase evolutions are as for the rays 11 and 21, respectively. New situation appears for the ray 32.

\paragraph{Ray 32:\hspace{2mm} $\pmb{ 0< \tilde c < \mu_{2}/\mu_{1} } $.}
The Universe started from the $EWs$ state. Then at the temperature given by \eqref{TEWSBI_2}
it went through the EWSB second-order phase transition into the inert-like phase $I_2$.
At the subsequent cooling down the Universe goes through the mixed phase $M$ into
the present inert phase  $I_1$. The second-order phase transitions $I_2\to M$ and $M\to I_1$
happened at the following temperatures
\be
T_{phtr}:\qquad  T_{2,M}= \sqrt{\fr{\mu_1-R\mu_2}{\tilde c_1 - R\tilde c_2}},\qquad
  T_{M,1}= \sqrt{\fr{R\mu_1-\mu_2}{R\tilde c_1 - \tilde c_2}}
  \,.\label{tempIN}
\ee
In accordance with equations in sect.~\ref{secvacst}, at the transition point
$I_2\to M$  masses of $S_H$ and $h$ vanish, while at the transition point
$M\to I_1$ masses of $h$ and $D_H$ become 0. At small distance from
the transition point with temperature $T_{phtr}$ these masses grow as a function of the temperature $T$ as $M_a^2=A_a|T^2-T^2_{phtr}|$, with
different coefficients $A_a$.

\subsection{The case $\pmb{0>R>-1}$}\label{secevolc}

\begin{figure}[htb]
\centering
\includegraphics[width=0.5\textwidth]{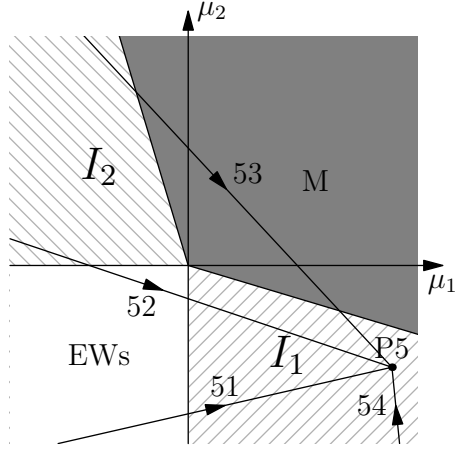}
\label{muplot3}
\caption{Phase diagram for $0>R>-1$ case. }
\end{figure}
The phase diagram is presented in Fig.~\ref{muplot3}.
In this case, as follows from  \eqref{Ccond2}, the  mixed phase $M$ region is realized
in a wider region than in Fig.~\ref{muplot2} (for $0<R<1$), even beyond
an upper right quadrant of this plane\fn{This case  was overlooked in the literature, see e.g. \cite{inert} and also \cite{limpap}. We thank G.~Gil and B.~Gorczyca for discussion on this point. }, namely:
\be
  \mu_2 > \mu_1/R ,\quad \mu_2 > \mu_1R.
    \label{Mphasec}
\ee
Since currently we are in the inert  vacuum, we have
\be
 \mu_2< R\mu_1\qquad (\mu_2<0).\label{inmuM}
\ee
Therefore, in this case we have only one type of today's point $P5$.
However, new opportunities appear due to larger freedom for temperature coefficients $c_i$, as  in accordance with  \eqref{signsci} in this case
$\tilde c_1$ and $\tilde c_2$ can be either positive
or negative.

 All possible phase evolutions leading to the point $P5$ are represented in
Fig.~\ref{muplot3} by four  rays 51, 52, 53 and 54.

The  { ray 51} describes similar evolution as rays 21 and 41.
New are rays 52, 53 and 54 with  common feature, which is
a lack of electroweak symmetry in very early stages of the Universe.

\paragraph{Ray 52:\hspace{2mm} $\pmb{\tilde c_1>0,\;\; \tilde c_2<0,\;\; \tilde c > \mu_2/\mu_{1}}$.}
Here a high-temperature  state of the
Universe is the inert-like vacuum $I_2$. With cooling down the Universe goes
through electroweak symmetric phase $EWs$ into the present $I_1$ phase. The second-order phase
transitions $I_2\to EWs$ and $EWs\to I_1$ happened, respectively,  at the
temperatures
\be
  T_{2,EWs}=\sqrt{{\mu_{2}}/{\tilde c_2}}\,,\qquad T_{EWs,1} =\sqrt{{\mu_{1}}/{\tilde c_1}}\,
.\label{ray52}
\ee

\paragraph{Ray 53:\hspace{2mm} $\pmb{\tilde c_1>0, \,\,\tilde c_2<0,\;\; \tilde c <\mu_{2}/\mu_{1} }$.}
Here a high-temperature state of the Universe is an inert-like
vacuum $I_2$. With cooling down the Universe passes through the mixed phase $M$ into
the present $I_1$ phase. The phase transitions $I_2\to M$ and $M\to I_1$ are of the
second order; they happened at the temperatures given by eqs.~\eqref{tempIN}.

\paragraph{Ray 54:\hspace{2mm} $\pmb{ \tilde c_1<0, \,\,\tilde c_2>0}$.}
For this ray the Universe stays in the inert vacuum $I_1$ during the whole
 evolution.

\subsection{Summary of possible evolutions}

We find that in the considered approximation the thermal evolution of Universe to the current  inert phase
can be studied effectively in the ($\mu_1,\mu_2$) plane,  at fixed values of quartic parameters $\lambda_i$. Different types of such  evolution represented as directed rays  depend crucially on two parameters: $R$~\eqref{lamnot}, describing the allowed for various vacua regions of the ($\mu_1,\mu_2$)
plane, and $\tilde{c}$~\eqref{ctilde}, determining
the direction of rays. The first one depends only on the ratios between coefficients of quartic part of potential $\lambda_i$, while the second depends both on mentioned parameters $\lambda_i$,
which are unknown up to now, and on precisely known gauge and Yukawa couplings.

One can distinguish following types of evolution to the current inert phase:
\begin{itemize}
\item{I.} The Universe evolves from the initial electroweak  symmetric state.
\begin{itemize}
\item{Ia.} The simplest evolution to the inert phase is realized through
a single EWSB phase transition of the second-order type (rays 11, 21, 31, 41, 51). Dark matter appears at this single transition simultaneously with EWSB.
\item{Ib.}  After the first EWSB phase transition Universe passes into the inert-like phase $I_2$. Then it passes  into the inert phase $I_1$ either directly (first-order phase transition, ray 12), or through  the  mixed phase $M$ (two second-order phase transitions, ray 32). In both cases dark matter appears only after the last phase transition to the inert phase.
\end{itemize}
\item{II.} The Universe evolves from the initial state having no electroweak
symmetry\fn{Such opportunity was discussed by a number of authors -- see e.g. \cite{nonEW}. Certainly, it is not ruled out, but it contradicts a key idea of modern approach -- the  state at very high temperatures  has high symmetry which is broken at cooling down of the Universe. In this sense this opportunity is unnatural.}.
\begin{itemize}
\item{IIa.} The initial phase is the inert one $I_1$. Evolution contains no phase transitions. Dark matter existed always (ray 54).
\item{IIb.} The initial phase is the inert-like one $I_2$. It contains no dark matter. Evolution to the current inert phase $I_1$ undergoes through two second-order phase transitions either via the mixed phase $M$ (ray 53), or via the  EWs phase, i.e. with a {\it temporary} appearance of electroweak  symmetry (ray 52). In both cases dark matter appears only after the last phase transition to the inert phase $I_1$.
\end{itemize}
\end{itemize}
Each of these evolutions can be realized in wide range of parameters.

\section{If DM is charged}\label{secchDM}

Model independent analysis shows that the case with charged DM particle is not ruled out absolutely, but charged DM particles must be heavier than $100q$~TeV, where $q$ is electric charge of DM particle in units of electron charge
\cite{chargedD}. For IDM it means $M_{H^\pm}>100$~TeV.
Such a heavy mass seems to be  unnatural in the  modern
particle physics  with natural energy scale   $\lesssim 1$~TeV. In this case, our new model will include both new particles
and new energy scale of phenomena.
Obviously, such an opportunity cannot be tested at colliders  in the estimable future.

The  case  with  charged  DM  particle  can be realized in IDM only if
$\lambda_4+\lambda_5>0$ \eqref{massesA}.
Since unitarity and perturbativity constraints
$|\lambda_i|\lesssim 8\pi$ must hold (see for details \cite{GIv,pert,GK05})
very large $D^\pm$ mass can arise only from
very large negative $m^2_{22}$ ~\eqref{massesA}. The position on the $(\mu_1,\mu_2)$ plane  of the actual state of the
Universe, for the anticipated in the SM value of the Higgs mass  $M_{h_S}\lesssim 200$~GeV and $M_D\pm\ge 100$~TeV,
corresponds to $\mu_1>0, \mu_2<0$, with large ratio of their absolute values
$\gtrsim 10^5-10^6$, if $\lambda_1/\lambda_2 \sim 1$.

In accordance with results of refs.~\cite{GK07,lorenzo,dsthesis}, if $\lambda_4+ \lambda_5>0$ then the mixed phase $M$ cannot exist (see e.g.~\eqref{chneitr}) while the charge breaking phase  can. The charge breaking vacuum can be realized if in addition $|R_3|<1$, where $R_3=\lambda_3/\sqrt{\lambda_1\lambda_2}$. Simple analysis shows that the phase diagrams for this case are similar to those in Figs.~\ref{muplot2},~\ref{muplot3}, with the replacement  $R\to R_3$. Rays  similar  to the rays 41, 51, 52 and 54 give nothing new in comparison with the cases discussed  in sect.~\ref{secevol}.

The really new opportunity could appear for the ray similar to ray 53 in Fig.~\ref{muplot3}
(with rays  going  through the charge breaking vacuum). However, this opportunity is ruled out. Indeed, to realize it one needs
$c_2<0$ and $|c_2|/c_1>|m_{22}^2|/m_{11}^2\gtrsim (10^5\div 10^6)$. The latter inequality contradicts
the $c_1>-c_2$ relation \eqref{cesum} based on the positivity condition.
It means that in our simple model   Universe evolution to the current inert phase  cannot pass through  the charge breaking phase.

\section{Results and discussion}

\paragraph{Main results.} The most important observation we made in this paper is as follows:
if current state of the Universe is described by IDM, then during the thermal evolution
the Universe can pass through various intermediate phases, different from the inert one.
 These possible intermediate phases contain no  dark matter, which
 appears only at the relatively late stage of  cooling down of the Universe.

A complete set of possible ways of evolution of the Universe,  including both EW symmetric ($EWs$)
 and EW non-symmetric (EW violating, $EWv$) initial states,
 can be summarized as follows\fn{Symbol I or II over arrows corresponds to the type of phase transition}:

\bear{c}
  \boxed{ EWs\xrightarrow{II}\left\{\begin{matrix}I_1&\,\hspace{1cm}\,\,\,\,\,\,\,\,\,rays\; 11,\,21,\,31,\,41,\,51\\
   I_2&\left\{\begin{matrix}\xrightarrow{II}M&\xrightarrow{II}I_1&ray\, 32\\
   \xrightarrow{I}I_1\qquad&\,\;\;\,\,\,\,\,\,\,\,ray\; 12\;&\end{matrix}\right.\end{matrix}\right.}\\[8mm]
 \boxed{ EWv:\qquad\begin{matrix}  I_2\xrightarrow{II}\left\{\begin{matrix}EWs \xrightarrow{II}I_1&ray\;52\\
   M \xrightarrow{II}I_1&ray\;53 \end{matrix}\right.\\
   I_1\to I_1 \hspace{23.5mm} ray\; 54\end{matrix}}
\eear{ways}

We see that both a simple EWSB with a direct transition to
the inert phase, as well as  sequences of two or three transitions from $EWs$ to the inert phase are possible. We found also that the current  inert state of Universe can be obtained from both initial high temperature state with EW symmetry and from the initial state without this symmetry.

As far the charged DM is concerned, still in principle allowed by the data, if heavy enough,  it seems to be excluded in IDM.

To find what scenario of evolution is realized in nature, one should measure all parameters of potential. The program how to measure these parameters at LHC and ILC is under preparation.

\paragraph{Outlook.} In contrast to the standard picture,
these scenarios allow for  the phase transition to the current inert phase  at relatively low  temperature,
giving new starting point for calculation of a today's abundance of the neutral DM components of the Universe.

In  this  paper  we  calculated  thermal  evolution of the Universe in the very high temperature
 approximation, i.e. for $T^2\gg |m_{ii}^2|$.
 The most interesting effects are expected at lower temperatures,
 where more precise calculations are necessary.  The  simplest expected modifications  of the presented
 description are:
\begin{enumerate}
\item Appearance of cubic terms like $\phi^3 T$  \cite{phi3}. These terms are important near
phase transition point, as they can transform some second-order phase
transitions
into the first-order transitions.
 \item The parameters become depend on temperature in more complicated way  than that  given by \eqref{Tempdep}. Therefore, the rays, depicted  thermal evolutions in Figs.~\ref{muplot1},~\ref{muplot2} and~\ref{muplot3}, can become non-straight. The bending of these rays can be different in different points of our plots and at different $\lambda_i$. It can give possible spectrum of phase evolutions even reacher that discussed above.
\end{enumerate}
However, we expect that the general picture will not change too much.

\paragraph{Possible extensions of model.}
The model we focused on in this paper contains two doublets. One can consider similar model with three doublets (particular case of 3HDM),
as it is discussed in ref.~\cite{Bogdan}.
It contains two standard Higgs doublets of 2HDM $\phi_{S1}$ and $\phi_{S2}$,
coupled to fermions, and one Higgs doublet $\phi_D$ (having no coupling to
fermions). The  potential is
invariant under $S, \,\,D$ transformations:
\bear{c}
    \{\phi_{S1},\,\phi_{S2}\} \xrightarrow{S} -\{\phi_{S1},\,\phi_{S2}\},\quad
   \phi_D \xrightarrow{S} \phi_D,\quad
   SM     \xrightarrow{S} SM; \\
   \{\phi_{S1},\,\phi_{S2}\} \xrightarrow{D} \{\phi_{S1},\,\phi_{S2}\},\quad
   \phi_D \xrightarrow{D} -\phi_D,
   SM     \xrightarrow{D} SM.\\[2mm]
\eear{DSinvBogd}

This model can incorporate all phenomena which appear in the standard 2HDM, at the same time it contains DM particles as in the IDM. Thermal evolution of its parameters  gives very diverse
phase diagram which contain in addition to the phases discussed above other phases,
discussed in refs.~\cite{GK07,GIK09,Ivanov:2008er} and  their mixtures.
However, even for this model our main conclusion about possible transformation
of the Universe through the phase without DM holds.

Reacher variants of both $S$- and $D$-sectors can be considered
similarly.  For example, one widely  discussed model of this type
(see e.g. \cite{Singledoubl})
contains the same $S$-sector as in  our paper,
but $D$-sector consists of one doublet and one singlet scalars,
non-interacting with fermions. This model has phenomenology
similar to that discussed above but one  hope to derive more or
less natural values of couplings starting from SO(10) universality at the GUT scale.
The biography of Universe  in this model can be studied as above --
obviously it will give more diverse phase story.

\section*{Acknowledgement}

We are thankful to I. Ivanov, M. Dubinin and R. Nevzorov for useful discussions. MK and DS would like to thank Grzegorz Gil and Bogus\l awa Gorczyca for important clarification, as well as to Piotr Chankowski for useful comments. Work was partly supported by Polish Ministry of Science and Higher Education Grant N N202 230337. The work of MK and DS was supported in addition  by  EU Marie Curie Research Training Network HEPTOOLS, under contract MRTN-CT-2006-035505, FLAVIAnet contract No. MRTN-CT-2006-035482.
The work of IG and KK was also supported by grants RFBR 08-02-00334-a, NSh-3810.2010.2 and Program of Dept. of Phys. Sc. RAS "Experimental and theoretical studies of fundamental interactions related to LHC."

\end{document}